\documentclass[12pt]{article}
\usepackage{graphicx}
\usepackage[cp1251]{inputenc}
\usepackage{epsfig}
\usepackage[english]{babel}

\date{}

\begin{document}
\setcounter{page}{1}
\pagestyle{plain}

\title{\bf{Electronic properties of armchair AA-stacked bilayer graphene nanoribbons}}

\author{Yawar Mohammadi$^1$\thanks{Corresponding author. Tel./fax: +98 831 427
4569, Tel: +98 831 427 4569. E-mail address:
yawar.mohammadi@gmail.com}  and Borhan Arghavani Nia$^2$}
\maketitle{\centerline{$^1$Young Researchers and Elite Club,
Kermanshah Branch, Islamic Azad University, Kermanshah,
Iran}\maketitle{\centerline{$^2$Department of Physics, Islamic
Azad University, Kermanshah Branch, Kermanshah, Iran}

\begin{abstract}

We study analytically, based on the tight-binding model, the
electronic band structure of armchair AA-stacked bilayer graphene
nanoribbons (BLGNRs) in several regimes. We apply hard-wall
boundary conditions to determine the discretion dominating on the
Bloch wavefunctions in the confined direction. First we consider
an ideal case, perfect nanoribbons without any edge deformation,
and show that their electronic properties are strongly
size-dependent. We find that the narrow armchair AA-stacked BLGNRs
(similar to single-layer graphene nanoribbons) may be metallic or
semiconducting depending on their width determined by the number
of dimer lines across the ribbon width, while the wide ribbons are
metallic. Then we show that, when the edge deformation effects are
taken into account, all narrow armchair AA-stacked BLGNRs become
semiconducting while the wide ribbons remain metallic. We also
investigate effects of an electric filed applied perpendicular to
the nanoribbon layers and show it can be used to tune the
electronic properties of these nanoribbons leading to a
semiconducting-to-metallic phase transition at a critical value of
the electric field which depends on the nanoribbon width.
Furthermore, in all regimes, we calculate the corresponding
wavefunctions which can be used to investigate and predict various
properties in these nanoribbons.
\end{abstract}


\vspace{0.5cm}

\section{Introduction}
\label{sec:01}

Graphene, an isolated single layer of graphite, since its
isolation in 2004~\cite{Novoselov01} has attracted many
experimental and theoretical research activities, leading to
discovery of many interesting properties~\cite{CastroNeto02} not
been observed in the ordinary two dimensional electron gases.
These unusual properties originate from the linear dependence of
its spectrum and the chiral nature of its quasiparticles. The
AB-stacked bilayer, and other few-layer graphene lattices,
inheriting the chiral nature from graphene, also exhibit
interesting properties~\cite{CastroNeto02,Peres03,DasSarma04}.
Recently a new stable stacking order of few-layer graphene,
few-layer graphene lattices with AA stacking order, has been
observed in experimental researches~\cite{Liu05,Borysiuk06}. In
these few-layer graphene lattices, each carbon atom in a top layer
is located directly above the same one in the bottom layer,
leading to special band structure composed of electron-, hole- and
even un-doped linear graphene band structures~\cite{Borysiuk06}.
The bilayer graphene with AA stacking order, due to this special
band structure, shows many interesting properties
~\cite{Ando07,Hsu08,Prada09,Rakhmanov10,Tabert11,Brey12} not been
reported for other materials.

The various experimental methods such as tailoring via a scanning
tunnelling microscopy tip~\cite{Hiura13}, exfoliating from highly
oriented pyrolytic graphite~\cite{Novoselov14,Zhang15,Bunch16} and
graphitizating SiC wafers~\cite{Berger17} can be used to fabricate
ribbons with finite width from single-layer graphene and from
other few-layer graphene. The single-layer graphene ribbons
(SLGNRs) have different width and different atomic edge
terminations. Studying effects of the size and the geometry, which
are found to be important in the nanotubes~\cite{Saito18}, is also
necessary for the nanoribbons of single- and few-layer graphene.

The graphene nanoribbons, depending on their edge shape, are
separated into two groups, armchair and zigzag graphene
nanoribbons. Based on the tight-binding model investigations, all
zigzag SLGNRs are metallic~\cite{Ezawa19,Brey20}. While all
armchair SLGNRs, due to the edge deformation effects and the
quantum confinement coming from the finite width of the
nanoribbon, are semiconducting~\cite{Rozhkov21,Zheng22}. A density
functional theory calculation~\cite{Son23}, which was confirmed
later by an experimental research~\cite{Han24}, showed that not
only all armchair SLGNRs but also all zigzag SLGNRs are
semiconducting. Recently AA-stacked BLG nanoribbons also have been
investigated~\cite{Habib25,Xu26,Zhong27,Ahmed28,Morell29}. In this
paper we derive, by taking the edge deformations into account, a
general analytical expression for the electronic band structure
and the wavefunction of the armchair AA-stacked BLGNRs which can
be used to investigate and predict various properties in these
nanoribbons. We also show that it is possible to tune the
electronic properties of the armchair AA-stacked BLGNRs via an
electric field applied perpendicular to the ribbons layers.

The paper is organized as follow. In the section II, we obtain
sublattice Bloch wavefunctions by imposing the hard-wall
condition. Then, by making use of the tight-binding model, we
present a general expression for the Schr\"{o}dinger equation of
the armchair AA-stacked BLGNRs which can be solved to obtain its
dispersion relation. First we consider an ideal case, an armchair
AA-stacked BLGNR without any edge deformation. Then we investigate
effects of the edge deformation on the dispersion relation and the
wavefunction of the armchair AA-stacked BLGNRs. In the section
III, we examine effects of an electric filed, applied
perpendicular to the nanoribbons layers, on the electronic
properties of the AA-stacked BLGNR. Finally, in the section IV, we
end the paper by summary and conclusions.

\section{Model Hamiltonian, electron wavefunction and band structure}
\label{sec:02}

In an AA-stacked BLGNR, which is composed of two SLG nanoribbons,
each sublattice in the top layer is located directly above the
same one in the bottom layer. Figure \ref{fig:01} shows a segment
of an AA-stacked BLGNR with armchair atomic edge termination whose
unit cell is enclosed inside the dashed lines. The unit cell
contains N A-type atoms and N B-type atoms in each layer that N is
the number of dimer lines across the ribbon width.

To investigate the electronic properties of a lattice structure
using tight-binding model, first we must construct its Bloch
wavefunction which for an AA-stacked BLGNR can be written as
\begin{eqnarray}
|\Psi\rangle=c_{A_{1}}|\psi_{A_{1}}\rangle+
c_{B_{1}}|\psi_{B_{1}}\rangle+c_{A_{2}}|
\psi_{A_{2}}\rangle+c_{B_{2}}|\psi_{B_{2}}\rangle, \label{eq:01}
\end{eqnarray}
where $|\psi_{A_{1}}\rangle$, $|\psi_{B_{1}}\rangle$,
$|\psi_{A_{2}}\rangle$ and $|\psi_{B_{2}}\rangle$ are the Bloch
wavefunction corresponding to $p^{z}$ orbital of $A_{1}$, $B_{1}$,
$A_{2}$ and $B_{2}$ sublattices respectively and $c_{A_{1}}$,
$c_{B_{1}}$, $c_{A_{2}}$ and $c_{B_{2}}$ are momentum-dependent
coefficients.

The armchair AA-stacked BLGN is supposed to has infinite length
(fig. \ref{fig:01}) along the x direction. This leads to
translational invariance along the x direction,
 allowing us to choose
plan wave basis along the x direction, $e^{ik_{x}x}$, for
constructing sublattice Bloch wavefunctions. While the quantum
confinement, coming from the finite width of the nanoribbon,
breaks the spectrum of an AA-stacked BLG into a set of subbands
and dominates a discretion on the amount of the momentum in the y
direction. The discrete values of the momentum in the y direction,
$k_{n}$, can be obtained by imposing hard-wall boundary condition
which indicates that all sublattice Bloch wavefunctions at
auxiliary sublattices at the both edges of the armchair AA-stacked
BLGN, $y=0$ and $y=(N+1)a/2$, must be zero. This give rises to
\begin{eqnarray}
k_{n}=n\frac{\pi}{N+1}\frac{2}{\sqrt{3}a},& & n=1,2,...N.
\label{eq:02}
\end{eqnarray}
Therefore, the normalized sublattice Bloch wavefunctions can be
written as
\begin{eqnarray}
|\psi_{A_{1}}\rangle=\frac{\sqrt{2}}{\sqrt{N_{x}(N+1)}}
\sum_{x_{A_{1}i}}\sum_{i=1}^{N}e^{ik_{x}x_{A_{1i}}}
\sin(\frac{\sqrt{3}a k_{n}}{2}i)|p^{z}_{A_{1i}}\rangle , \nonumber
\\
|\psi_{B_{1}}\rangle=\frac{\sqrt{2}}{\sqrt{N_{x}(N+1)}}
\sum_{x_{B_{1}i}}\sum_{i=1}^{N}e^{ik_{x}x_{B_{1i}}}
\sin(\frac{\sqrt{3}a k_{n}}{2}i)|p^{z}_{B_{1i}}\rangle , \nonumber
\\
|\psi_{A_{2}}\rangle=\frac{\sqrt{2}}{\sqrt{N_{x}(N+1)}}
\sum_{x_{A_{2}i}}\sum_{i=1}^{N}e^{ik_{x}x_{A_{2i}}}
\sin(\frac{\sqrt{3}a k_{n}}{2}i)|p^{z}_{A_{2i}}\rangle , \nonumber
\\
|\psi_{B_{2}}\rangle=\frac{\sqrt{2}}{\sqrt{N_{x}(N+1)}}
\sum_{x_{B_{2}i}}\sum_{i=1}^{N}e^{ik_{x}x_{B_{2i}}}
\sin(\frac{\sqrt{3}a k_{n}}{2}i)|p^{z}_{B_{2i}}\rangle ,
\label{eq:03}
\end{eqnarray}
where $N_{x}$ is the number of unit cells along the x direction
and $x_{A_{1i}}$ is the x-coordinate of $A_{1}$ in the i-th dimer
line across the ribbon width. The spectrum and the wavefunction
are derived by solving the Schr\"{o}dinger
equation~\cite{Grosso30} which, for an armchair AA-stacked BLGNR,
reduces to a $4\times4$ matrix equation as
\begin{eqnarray}
\left(
\begin{array}{cccc}
H_{A_{1}A_{1}}-E &  H_{A_{1}B_{1}} & H_{A_{1}A_{2}} & H_{A_{1}B_{2}} \\
H_{B_{1}A_{1}} & H_{B_{1}B_{1}}-E & H_{B_{1}A_{2}} & H_{B_{1}B_{2}} \\
H_{A_{2}A_{1}} & H_{A_{2}B_{1}} & H_{A_{2}A_{2}}-E & H_{A_{2}B_{2}} \\
H_{B_{2}A_{1}} & H_{B_{2}B_{1}} & H_{B_{2}A_{2}} &
H_{B_{2}B_{2}}-E
\end{array}
\right)
 \left(
\begin{array}{c}
c_{A_{1}} \\
c_{B_{1}} \\
c_{A_{2}} \\
c_{B_{2}}
\end{array}
\right)=0, \label{eq:04}
\end{eqnarray}
where
$H_{A_{p}A_{q}}=\langle\psi_{A_{p}}|\mathcal{H}|\psi_{A_{q}}\rangle$,
$H_{B_{p}B_{q}}=\langle\psi_{B_{p}}|\mathcal{H}|\psi_{B_{q}}\rangle$
and
$H_{A_{p}B_{q}}=H^{\ast}_{B_{q}A_{p}}=\langle\psi_{A_{p}}|\mathcal{H}|\psi_{B_{q}}\rangle$
where $p$ and $q$ are $1$ and $2$. Here, $\mathcal{H}$ is the
Hamiltonian of an armchair AA-stacked BLGNR lattice.

\subsection{Perfect nanaoribbons}

First we consider an ideal case, a perfect armchair AA-stacked
BLGNR \textit{without any edge deformation}. If we apply the
nearest-neighbor tight-binding approximation to obtain the
elements of the matrix equation (Eq. \ref{eq:04})), we get
\begin{eqnarray}
H^{0}_{A_{1}B_{1}}=H^{0\ast}_{B_{1}A_{1}}=H^{0}_{A_{2}B_{2}}=H^{0\ast}_{B_{2}A_{2}}=f^{0}_{n}(k_{x}),
\nonumber
\\
H^{0}_{A_{1}A_{1}}=H^{0}_{B_{1}B_{1}}=H^{0}_{A_{2}A_{2}}=H^{0}_{B_{2}B_{2}}=\varepsilon_{p^{z}}
, \nonumber
\\
H^{0}_{A_{1}A_{2}}=H^{0}_{A_{2}A_{1}}=H^{0}_{B_{1}B_{2}}=H^{0}_{B_{2}B_{1}}=\gamma,
\nonumber
\\
H^{0}_{A_{1}B_{2}}=H^{0}_{B_{1}A_{2}}=H^{0}_{A_{2}B_{1}}=H^{0}_{B_{2}A_{1}}=0,
\label{eq:05}
\end{eqnarray}
where
$f^{0}_{n}(k_{x})=-t(e^{-ik_{x}a}+2\cos(n\frac{\pi}{N+1})e^{ik_{x}a/2})$,
$t\sim 3 eV$~\cite{Tabert11} and $\gamma\sim0.2
eV$~\cite{Xu31,Lobato32} are the nearest-neighbor intralyer and
interlayer hopping energies and $\varepsilon_{p^{z}}$ is energy of
the $p^{z}$ orbital which can be kept to zero as the energy
reference. In the above equations and hereafter, the superscript
$0$ indicates that these quantities refer to a perfect armchair
AA-stacked BLGNR, an armchair AA-stacked BLGNR ribbon without any
edge deformation. Here, we have used
$S_{A_{p}A_{p}}=\langle\psi_{A_{p}}|\psi_{A_{p}}\rangle=1$,
$S_{B_{p}B_{p}}=\langle\psi_{B_{p}}|\psi_{B_{p}}\rangle=1$ and
$S_{A_{p}B_{q}}=S^{\ast}_{B_{q}A_{p}}=\langle\psi_{A_{p}}|\psi_{B_{q}}\rangle\approx0$~\cite{Reich33}
where $p$ and $q$ are $1$ and $2$. It is the so-called orthogonal
tight-binding schemes. Thus, the spectrum and the wavefunction
become
\begin{eqnarray}
&&E_{n}^{0s\lambda}=s\gamma+\lambda|f^{0}_{n}(k_{x})| , \nonumber
\\
&&|\Psi\rangle^{0s\lambda}_{n}=\frac{1}{2}([|\psi_{A_{1}}\rangle+\lambda
\frac{f^{0\ast}_{n}(k_{x})}{|f^{0}_{n}(k_{x})|}|\psi_{B_{1}}\rangle]
+s[|\psi_{A_{2}}\rangle
+\lambda\frac{f^{0\ast}_{n}(k_{x})}{|f^{0}_{n}(k_{x})|}|\psi_{B_{2}}\rangle]),
\label{eq:06}
\end{eqnarray}
where $s=\pm$ are band indexes and $\lambda=\pm$ denote the
conduction and the valance bands. It is evident from Eq.
(\ref{eq:06}) that the spectrum of an armchair AA-stacked BLGNR is
composed of the spectrums of two armchair SLGNRs denoted by $s=+$
and $s=-$ which have been shifted along energy axis by $+\gamma$
and $-\gamma$ respectively. This make the electronic band
structure of armchair AA-stacked BLGNRs different from that of
SLGNRs. Moreover, as Eq. (\ref{eq:06}) shows, since all $A_{1}$,
$B_{1}$, $A_{2}$, and $B_{2}$-atoms of each dimer contribute
equivalently in constructing wavefunction, the electron density of
sates on all atoms of each dimer are equal and the electron
density of states only depend on the distance from the armchair
edge. This can be examined by a STM image.

The armchair AA-stacked BLGNRs, depending on their width, may be
metallic or semiconducting. Figure \ref{fig:02} shows our results
for the energy gap of different armchair AA-stacked BLGNRs as a
function of their width. According to the width dependence of the
energy gap, the armchair AA-stacked BLGNRs are separated into
three groups denoted by $N=3m$, $N=3m+1$ and $N=3m+2$ where $N$ is
the number of dimer lines across the ribbon width (figure
\ref{fig:01}) and $m$ is a positive integer number. This figure
shows that all armchair AA-stacked BLGNRs with $N=3m+2$ are
metallic. This is understood by the fact that for these
nanoribbons,
$f^{0}_{n=2m+2}(k_{x}=0)=-2t(\cos(\frac{(2m+2)\pi}{3m+3})+\frac{1}{2})=2t(\cos(\frac{2\pi}{3})+\frac{1}{2})=0$.
This means that, for this group of armchair AA-stacked BLGNRs, the
upmost valance and the lowest conduction sub-bands of bands with
same s-index band always touch each other at $k_{x}=0$, leading to
metallic behavior.

Moreover, this figure shows that the armchair AA-stacked BLGNRs
with $N=3m$ and $N=3m+1$ may be metallic or semiconducting
depending on their width. This is different from what has been
reported for armchair
SLGNRs~\cite{Ezawa19,Brey20,Rozhkov21,Zheng22}. This can be
explained as follows; It is evident that for $N=3m$ and $N=3m+1$
all $f^{0}_{n}(k_{x}=0)\neq0$, inducing an energy gap between the
sub-bands with same s-index band. Moreover, as mentioned above,
the spectrum of an AA-stacked BLGNR is composed of two SLGNR
spectrums denoted by $s=+$ and $s=-$ which have been shifted along
energy axis by $+\gamma$ and $-\gamma$ respectively. Therefore, an
armchair AA-stacked BLGNR is semiconducting if the induced bad
gaps, the minimum of $2t|1+2\cos(\frac{n\pi}{N+1})|$ with
$n=1,2,...N$ which is $2t|1+2\cos(\frac{(2m+1)\pi}{N+1})|$, is
larger than $2\gamma$ and it is metallic when it isn't. This can
be seen in  fig. \ref{fig:03} and fig. \ref{fig:04}. This is
dependent on the width of the nanoribbon. When the nanoribbon is
narrow, the induced energy gap is large enough to be larger than
$2\gamma$, so the nanoribbon become semiconducting. While when the
width of the nanoribbon increases, the induced energy gap becomes
smaller than $2\gamma$ leading to the metallic behavior. Notice
that for both groups of semiconducting armchair AA-stacked BLGNRs
($N=3m$ and $N=3m+1$), there is a critical width that if the
nanoribbon width exceeds that the nanoribbon becomes metallic.

\subsection{Edge deformation effects}

The nanoribbons have unpaired covalent carbon bonds at their edges
making these material unstable~\cite{Rozhkov21}. These dangling
bonds can be passivated by hydrogen atoms or other kinds of atoms
or molecules~\cite{Ezawa19,Zheng22,Son23}, leading to a stable
state ~\cite{Rozhkov21}. The first-principle calculations showed
that the carbon bond-lengths at the edges of the hydrogenated
SLGNRs~\cite{Son23} and the large aromatic
molecules~\cite{Coulson34} are shorter than those in the middle
regions. Furthermore, an analytical calculation\cite{Porzag35},
based on the tight-binding model, found that a decreasing about 3
- 4 percent in the carbon bond-lengths could induce an increasing
about 12 percent in the hopping integral between $p^{z}$ orbitals.
These edge deformation effects must be taken into account to
achieve a real understanding about electronic properties of the
nanoribbons. Here, similar to the previous
works~\cite{Zheng22,Son23} which studied the armchair SLGNRs, we
suppose that all intralayer hopping energy deviations are
negligible except those occur between the sublattices at $i=1$
dimer line or between those at $i=N$ dimer line. Therefore, we
have
\begin{eqnarray}
\delta H_{A_{p}B_{p}}&=& \delta H^{\ast}_{B_{p}A_{p}}=\delta
f_{n}(k_{x})\nonumber
\\ &=&-\frac{2\delta
te^{-ik_{x}a}}{N+1}(\sin^{2}(\frac{n\pi}{N+1})+\sin^{2}(N\frac{n\pi}{N+1}))
\nonumber
\\ &= &-\frac{4\delta
te^{-ik_{x}a}}{N+1}\sin^{2}(\frac{n\pi}{N+1}),\label{eq:07}
\end{eqnarray}
where $\delta t$ is the nearest-neighbor intralayer hopping energy
deviation and $p=1$ and $2$. So, the spectrum and the wavefunction
of the edge-deformed armchair AA-stacked BLGNR are given by
\begin{eqnarray}
&&E_{n}^{s\lambda}=s\gamma+\lambda|f_{n}(k_{x})| , \nonumber
\\
&&|\Psi\rangle^{s\lambda}_{n}=
\frac{1}{2}([|\psi_{A_{1}}\rangle+\lambda
\frac{f^{\ast}_{n}(k_{x})}{|f_{n}(k_{x})|}|\psi_{B_{1}}\rangle]
+s[|\psi_{A_{2}}\rangle+\lambda\frac{f^{\ast}_{n}(k_{x})}{|f_{n}(k_{x})|}|\psi_{B_{2}}\rangle]),
\label{eq:08}
\end{eqnarray}
where $f_{n}(k_{x})=f^{0}_{n}(k_{x})+\delta f_{n}(k_{x})$ and so
\begin{eqnarray}
E_{n}^{s\lambda}=s\gamma+\lambda\sqrt{|f^{0}_{n}(k_{x})|^{2}+|\delta
f_{n}(k_{x})|^{2}+2Re(f^{0}_{n}(k_{x})\delta f^{\ast}_{n}(k_{x}))}
. \label{eq:09}
\end{eqnarray}
It is easy to show that the band gap of the deformed nanoribbon is
a direct one which occurs at $k_{x}=0$. The band gap occurs
between the first conduction and valance bands with $n=2m+1$ when
$N=3m$ or $N=3m+1$ and between the subbands with $n=2m+2$ when
$N=3m+2$. The analytical results for the energy gap, obtained from
Eq. (\ref{eq:09}) in terms of the $\delta t$ up to the first order
($|\delta f_{n}(k_{x})|^{2}\rightarrow0$), are given by
\begin{eqnarray}
\Delta^{BLG}_{N}=[\Delta^{SLG}_{N}-2\gamma
]\theta(\Delta^{SLG}_{N}-2\gamma),\label{eq:10}
\end{eqnarray}
where $\Delta^{SLG}_{N}$ are the energy gap of the armchair SLGNR
with same width which are
\begin{eqnarray}
&&\Delta^{SLG}_{N=3m}=2t|1+2\cos(\frac{2m+1}{N+1}\pi)|-\frac{8\delta
t}{N+1}\sin^{2}(\frac{m}{N+1}\pi),\nonumber
\\
&&\Delta^{SLG}_{N=3m+1}=2t|1+2\cos(\frac{2m+1}{N+1}\pi)|+\frac{8\delta
t}{N+1}\sin^{2}(\frac{m+1}{N+1}\pi), \nonumber
\\
&&\Delta^{SLG}_{N=3m+2}=\frac{6\delta t}{N+1}. \label{eq:11}
\end{eqnarray}
Our results at zero limit of the interlayer hopping energy reduce
to the results which have been reported for the armchair
SLGNRs~\cite{Zheng22,Son23}. Figure \ref{fig:05} shows our results
for the energy gap of the edge-deformed armchair AA-stacked BLGNRs
as a function of their width obtained from Eq. (\ref{eq:09}). We
see that, due to the edge deformation effects which couple
electron and hole states with equal momentum, a band gap opens in
the band structure of the narrow armchair AA-stacked BLGNRs with
$N=3m+2$. Furthermore, the energy gap of the the armchair
AA-stacked BLGNRs with $N=3m+1$ increases, while that of the
armchair AA-stacked BLGNRs with $N=3m$ decreases. Hence, all
narrow edge-deformed armchair AA-stacked BLGNRs become
semiconducting while the wide ribbons remain metallic. Moreover,
Eq. (\ref{eq:08}) shows that, when the edge deformation effects
are take into account, the magnitude of the wavefunction on all
sublattices of the edge-deformed armchair AA-stacked BLGNRs remain
unchanged indicating that the local density of state on all
sublattices of edge-deformed armchair AA-stacked BLGNRs are equal
to those of the perfect nanoribbons.

\section{Effects of a perpendicular electric field}

Here we investigate effects of a perpendicular electric field on
the electronic properties of armchair AA-stacked BLGNRs. Applying
a perpendicular electric field creates a potential $+V$ in the top
layer and a potential $-V$ in the bottom one. So the corresponding
Schr\"{o}dinger equation can be obtained from Eq. (\ref{eq:04}) by
just substituting

\begin{eqnarray}
H_{A_{1}A_{1}}\rightarrow H_{A_{1}A_{1}}+V &,&
H_{B_{1}B_{1}}\rightarrow H_{B_{1}B_{1}}+V,\nonumber \\
H_{A_{2}A_{2}}\rightarrow H_{A_{2}A_{2}}-V &,&
H_{B_{2}B_{2}}\rightarrow H_{B_{2}B_{2}}-V. \label{eq:12}
\end{eqnarray}

If we take effects of the edge deformation into account, this
yields
\begin{eqnarray}
&&E_{n}^{Vs\lambda}=s\gamma^{'}+\lambda|f_{n}(k_{x})| , \nonumber
\\
&&|\Psi\rangle^{Vs\lambda}_{n}=\frac{\gamma}{2\sqrt{\gamma^{'}(\gamma^{'}-V)}}([|\psi_{A_{1}}\rangle+\lambda
\frac{f^{\ast}_{n}(k_{x})}{|f_{n}(k_{x})|}|\psi_{B_{1}}\rangle] +s
\frac{\gamma^{'}-V}{\gamma}
[|\psi_{A_{2}}\rangle+\lambda\frac{f^{\ast}_{n}(k_{x})}{|f_{n}(k_{x})|}|\psi_{B_{2}}\rangle]),
\label{eq:13}
\end{eqnarray}
where $\gamma^{'}=\sqrt{{\gamma}^{2}+V^{2}}$. These results at the
zero limit of the electric potential, $V\rightarrow0$, reduce to
our result which has been introduced in Eq. (\ref{eq:08}). Notice
that effects of the perpendicular electric field on the electronic
band structure can be taken into account only by a renormalization
of the interlayer hoping energy to a new value which depends on
the electrical potential as
$\gamma^{'}=\sqrt{{\gamma}^{2}+V^{2}}$. This shows that one can
tune the electronic structure of the armchair AA-stacked BLGNRs by
an electric filed applied perpendicular to layers. Figure
\ref{fig:06} shows our results for the energy gap of the armchair
AA-stacked BLGNRs with $N=3m+1$ for different values of the
vertical electric field, $V=0.0$, $V=1.0\gamma$ and $V=2.0\gamma$.
Notice that if the amount of the electric field increase the
energy gap of the nanoribbon decreases leading to a
semiconducting-to-metallic phase transition at a special value of
the electric field. This indicates that these nanoribbons can be
used as current switchers. Moreover, this phase transition can
lead to other interesting properties, which have been reported for
undoped AA-stacked BLG coming from the nonzero density of state at
Fermi energy, such as coherent plasmon dispersion which can exist
even in the presence of doping\cite{Roldan36}. Based on our
results, in an armchair AA-stacked BLGNR this plasmon dispersion
can be controlled electrically.

As another result of Eq. (\ref{eq:08}), we mention that, in the
presence of a perpendicular electric field, the electron-density
of states at top and bottom layer becomes different. The reason is
that the perpendicular electric field breaks the symmetry of two
layers leading to different probability amplitudes on top and
bottom layer sublattices.

\section{Summary and conclusions}
\label{sec:03}

In summary, we derive analytical relations for the electronic band
structure and the wavefunction of AA-stacked BLGNRs with armchair
edge shapes using tight-binding model in several regimes. First we
considered an ideal case, armchair AA-stacked BLGNRs without any
edge deformation. We found that the electronic properties of these
nanoribbons, similar to armchair SLGNRs, depend on their width but
with several differences; All ribbons with $N=3m+2$ are metallic,
while for ribbons with $N=3m$ and $N=3m+1$, by increasing the
ribbon width a semiconducting-to-metallic phase transition takes
place, where $N$ is the number of dimer lines across the ribbon
width. In addition, we investigate edge deformation effects and
showed that all narrow edge-deformed armchair AA-stacked BLGNRs
become semiconducting while the wide ribbons remain metallic.
Moreover, we showed that our analytical results for the electron
wavefunction and the band structure at the limit of zero
interlayer hopping energy reduces to the results reported for the
armchair SLGNRs. Finally, we considered the effect of a
perpendicular electric field on the electronic properties of the
armchair AA-stacked BLGNRs. We showed that one can tune the
electronic band structure of the armchair AA-stacked BLGNRs
leading to some interesting properties which have been reported
for an AA-stacked BLG lattice.

%

%
%
\newpage
\begin{figure}
\begin{center}
\includegraphics[width=12cm,angle=0]{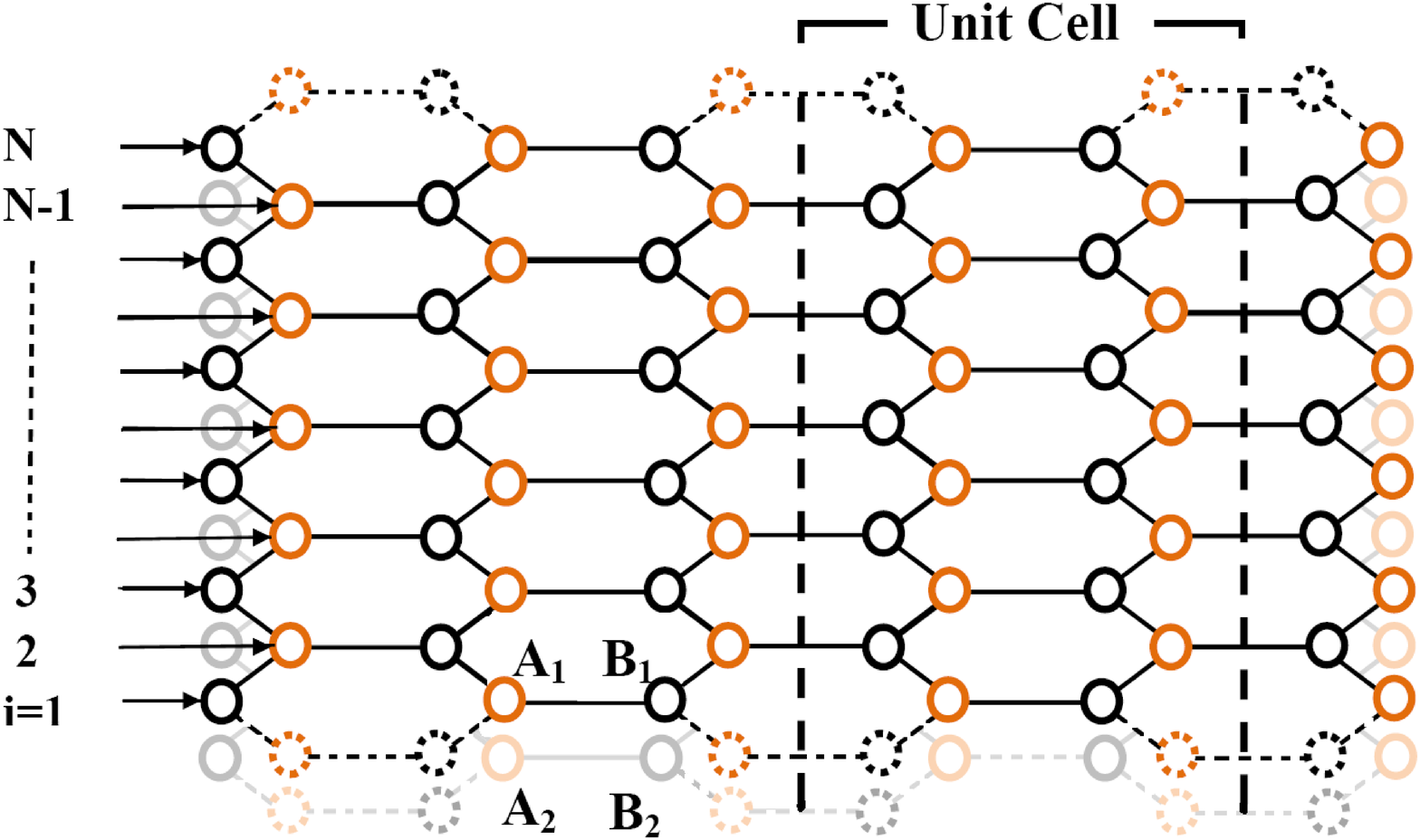}
\caption{Structure of an armchair AA-stacked BLGNR whose unit cell
is limited by two dashed lines, consisting $4N$ sublattices which
$N$ is the number of the dimer lines across its width. The dashed
circles at both edges of the ribbon denote auxiliary sublattices
where the sublattice wavefunctions are supposed to be zero there.
The sublattices in the top layer are indicated by subscript $1$,
$A_{1}$ and $B_{1}$, and those in the bottom layer by subindex
$2$, $A_{2}$ and $B_{2}$. }\label{fig:01}
\end{center}
\end{figure}
\begin{figure}
\begin{center}
\includegraphics[width=12cm,angle=0]{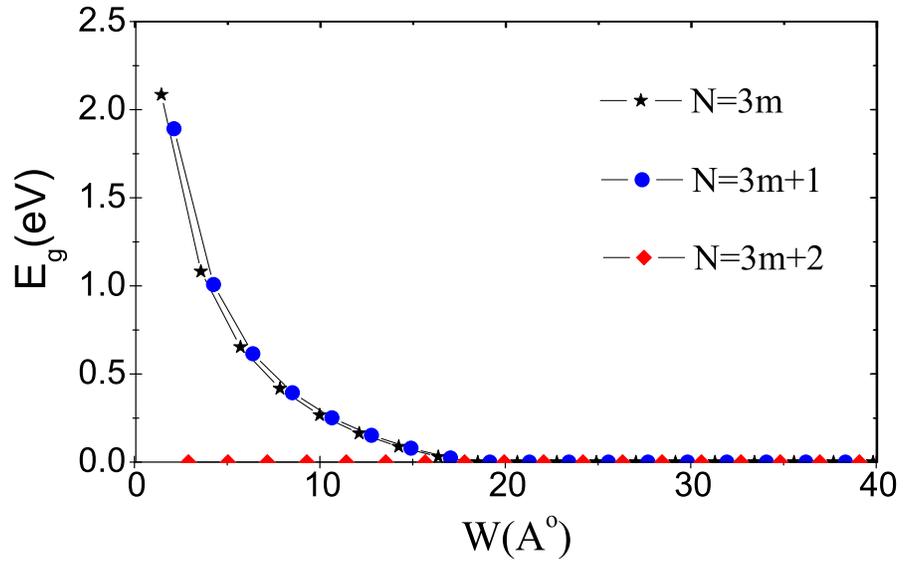}
\caption{The energy gap of the perfect armchair AA-stacked BLGNRs
as a function of their width. Here the edge deformation effects
have been neglected. Three different groups have been indicated by
the number of the dimer lines across their width, $N=3m$, $N=3m+1$
and $N=3m+2$. }\label{fig:02}
\end{center}
\end{figure}
\begin{figure}
\begin{center}
\includegraphics[width=12cm,angle=0]{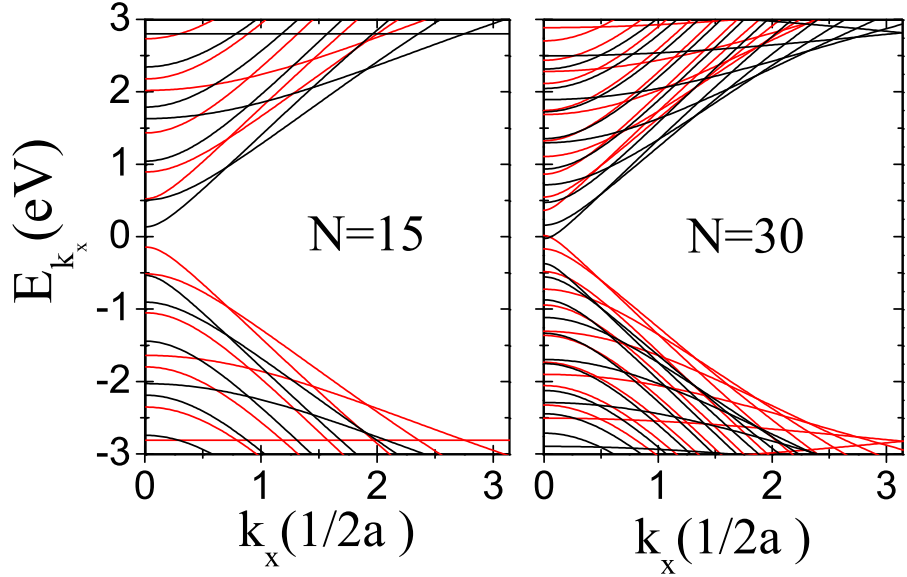}
\caption{The electronic band structure of two armchair AA-stacked
BLGNRs with $N=3m$, $N=15$ (left panel) and $N=30$ (right panel).
The red and black curves show the subbands with $s=+$ and $s=-$
respectively. Notice that for the armchair AA-stacked BLGNRs with
$N=30$, the induced band gap is filled by the shifted subbands,
leading to the metallic behavior. }\label{fig:03}
\end{center}
\end{figure}
\begin{figure}
\begin{center}
\includegraphics[width=12cm,angle=0]{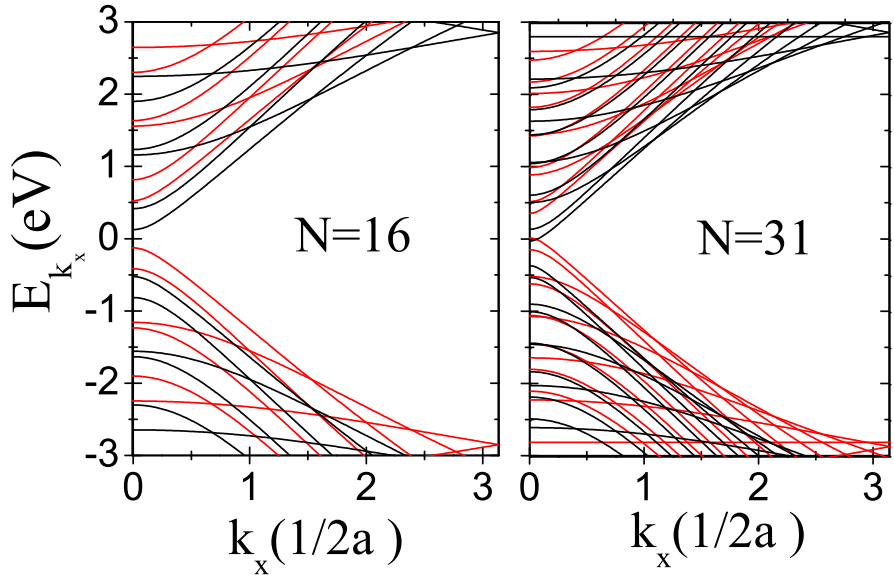}
\caption{Same as figure \ref{fig:03} but with $N=3m+1$.
}\label{fig:04}
\end{center}
\end{figure}
\begin{figure}
\begin{center}
\includegraphics[width=12cm,angle=0]{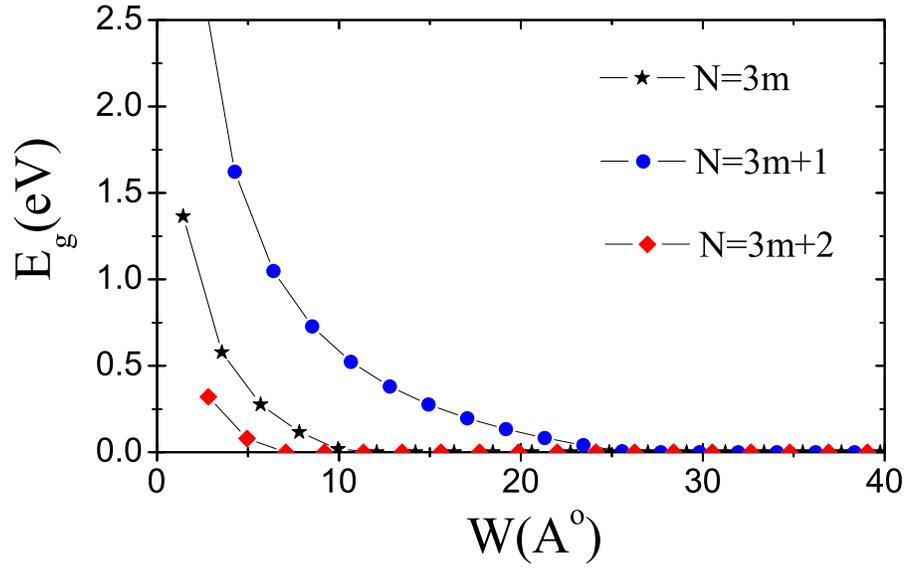}
\caption{The energy gap of the edge-deformed armchair AA-stacked
BLGNRs as a function of their width. Three different groups have
been indicated by the number of the dimer lines across their
width, $N=3m$, $N=3m+1$ and $N=3m+2$. }\label{fig:05}
\end{center}
\end{figure}
\begin{figure}
\begin{center}
\includegraphics[width=12cm,angle=0]{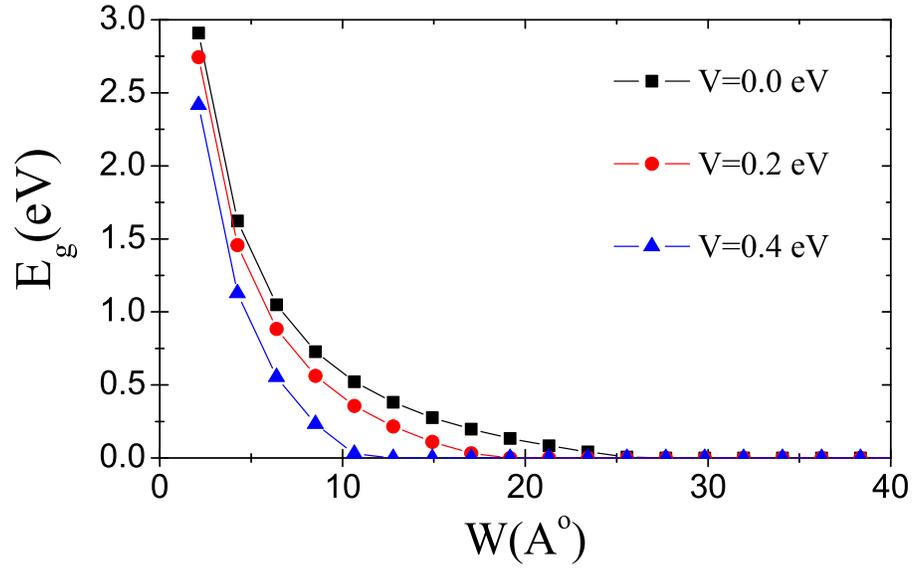}
\caption{The energy gap of the edge-deformed armchair AA-stacked
BLGNRs with $N=3m+1$ as a function of their width in the presence
of a vertical electric field, for three different values of the
electrical potential: $V=0.0 eV$, $V=0.2 eV$ and $V=0.4 eV$.
}\label{fig:06}
\end{center}
\end{figure}
\end{document}